\documentclass[aip, pre]{revtex4-1}

\usepackage{amssymb}
\usepackage{amsmath}
\usepackage{graphicx}
\usepackage{float}

\begin{document}

\title{Computational capabilities at the edge of chaos for one dimensional systems undergoing continuous transitions}

\author{E. \surname{Estevez-Rams}}
\email{estevez@fisica.uh.cu}
\affiliation{Facultad de F\'isica-Instituto de Ciencias y Tecnolog\'ia de Materiales(IMRE), University of Havana, San Lazaro y L. CP 10400. La Habana. Cuba.}

\author{D.  \surname{Estevez-Moya}}
\affiliation{Facultad de F\'isica, University of Havana, San Lazaro y L. CP 10400. La Habana. Cuba.}

\author{K.  \surname{Garcia-Medina}}
\affiliation{Facultad de F\'isica, University of Havana, San Lazaro y L. CP 10400. La Habana. Cuba.}

\author{R. \surname{Lora-Serrano}}
\affiliation{Universidade Federal de Uberlandia, AV. Joao Naves de Avila, 2121- Campus Santa Monica, CEP 38408-144, Minas Gerais, Brasil. }

\begin{abstract}
While there has been a keen interest in studying computation at the edge of chaos for dynamical systems undergoing a phase transition, this has come under question for cellular automata. We show that for continuously deformed cellular automata there is an enhancement of computation capabilities as the system moves towards cellular automata with chaotic spatiotemporal behavior. The computation capabilities are followed by looking into the Shannon entropy rate and the excess entropy, which allows identifying the balance between unpredictability and complexity. Enhanced computation power shows as an increase of excess entropy while the system entropy density has a sudden jump to values near one. The analysis is extended to a system of non-linear locally coupled oscillators that have been reported to exhibit spatiotemporal diagrams similar to cellular automata.
\end{abstract}


\date{\today}
\maketitle


\section{Introduction}

The transition from regular towards chaotic behavior is one of the defining characteristics of complex non-linear dynamical systems. There has been a keen interest in the behavior of a dynamical system as it undergoes such transition, which comes from the fact that it is precisely in such region where enhanced or improved computational capabilities of the system are claimed to be found \cite{crutchfield90,langton90,kauffman92}, this is known as the edge of chaos (EOC) hypothesis. In this context, computation in a dynamical system can have different meanings \cite{crutchfield93}. The most common interpretation of computational capacity is related to the ability of the system to perform some ``meaningful'' task. Meaningful can be understood, given some initial condition, as the system transforming the input in some prescribed way into some output data. More broadly, it can be understood related to the system capability to behave as a universal Turing machine given some proper input and ``programming''. A related but different point of view is to associate computation to the generic production, storage, transmission and manipulation of information. It is in this last sense that any dynamical system can be seen as some device amenable to be characterized by its intrinsic computational abilities \cite{crutchfield12}. Measures of the computational capabilities of a system have been developed that aim at quantifying the efficiency, in terms of resources, while characterizing the balance between irreducible randomness and pattern production \cite{crutchfield12}. 

In favor of the EOC hypothesis several systems modeling natural and artificial phenomena exhibit interesting, if not some type of critical behavior, precisely at the edge of chaos \cite{su89,sole96,bertschinger04,beggs08,boedecker12}. It is not clear, for some one dimensional systems, if the EOC criticality plays a role in the enhancement of computational capabilities, for example in cellular automata. Cellular automata (CA) are discrete space, dynamical systems that act over a spatially extended grid taking values over a finite alphabet, resulting in the evolution of the spatial configuration in discrete time steps. There has been a discussion if meaningful computation abilities in CA are found and should be sought at EOC regions. 

Langton \cite{langton90} has used the density of quiescent states in the cellular automaton rule, usually denoted by the letter $\lambda$, as the control parameter to explore such questions. The quiescent state is chosen as some arbitrary value in the CA alphabet, and the $\lambda$ parameter is the ratio between the number of times the quiescent state does not appear in the CA rule table and the number of entries in the same table. For a binary alphabet CA with nearest neighbor rules (elementary cellular automata or ECA), $\lambda=1-n\times2^ {-3}=1-n\times8$, where $n$ is the number of times the quiescent state (say value $1$) appears in the rule table. Monte Carlo simulations were performed over the CA space to conclude that, on average, as $\lambda$, increases from $0$, the rules behavior changes from a fixed point regime, to periodic, to complex and finally to chaotic, each regime corresponding to one of Wolfram classification scheme \cite{wolfram02}. Packard \cite{packard88} further developed these ideas to conclude that CA capable of meaningful computations happens precisely at those critical values of $\lambda$ where a transition to the chaotic behavior occurs. Packard results were questioned by Crutchfield et al. \cite{crutchfield93} who performed Monte-Carlo simulations similar to the ones made by Packard failing to find EOC enhanced computation capabilities.

In this paper, we show that improvement of computational capabilities indeed occurs at the edge of chaos for some continuously extended CA rules, but not following Langton $\lambda$ parameter and therefore, not in the sense of Packard simulations.  We will not be seeking to discover useful computational of the CA  but instead, improvement of intrinsic computation as measured by entropic based measures of structure and disorder. To support this claim we follow the behavior of a CA as one entry in its rule table is continuously varied between $0$ and $1$ and use it as control parameter the value of the continuously varying entry.  After discussing CA, we turn our attention to a system of one dimensional non-linear locally coupled oscillators that have been found to exhibit dynamical behavior similar to those of CA \cite{alonso17}. The quest is to find if in this system with a continuous control variable space, enhanced computation at the EOC can also be found.  

\section{Continuous states cellular automata}

To start formalizing some of the expressed ideas, for our purposes it suffices to consider one dimensional CA as a tuple $\{\mathcal{A}, r, \phi\}$ consisting of a set of states $\mathcal{A}$, a radius $r$ which defines a neighborhood structure from $-r$ to $r$, and $\phi$ a local rule $\phi: \mathcal{A}^{2r+1}\rightarrow \mathcal{A}$. The local rule acts simultaneously over a numerable ordered set $S$ of cells which can be bi-infinite, but in any case, each cell is properly labeled by natural numbers as in $\{\ldots, s_{-1}, s_0, s_1, \ldots\}$. At a given time step $t$ each cell is in a state belonging to $\mathcal{A}$ and the whole cell array configuration is denoted by $S^{(t)}$. The evolution of the cell array  from time $t$ to time $t+1$ is given by the global function induced by $\phi$: $\forall S \in \mathcal{A}^{\mathbb{Z}},\; S^{(t+1)}=\Phi(S^{t})$ such that $s_i^{(t+1)}=\phi(s_{i-r}^{(t)},\ldots, s_{i}^{(t)},\ldots,s_{i+r}^{(t)})$. $S^{(t)}$ is also called the spatial configuration of the cell array at time $t$. The set of $\{S^{(0)}, S^{(1)}, \ldots, S^{(t)}\}$ is called the spatiotemporal diagram or evolution up to time $t$. The spatiotemporal diagram of the CA can show a wide range of behaviors including the emergence of complex patterns, fractal nature, chaotic behavior, periodic, random evolution, or evolve to homogeneous spatial configurations.  CA have been used as models for different natural phenomena as well as a model for computing devices. It is known at least of a one dimensional CA that behaves as a Universal Turing Machine, the so-called rule 110 \cite{cook04,wolfram02}. 

Consider two state $\mathcal{A}=\{0,1\}$ cells with a rule defined over a one neighborhood radius $r=1$ such that, a cell $s_{i}$ will be updated in the next time step according to the current value of the cell and its two nearest neighbors: $s_{i}^{(t+1)}=\phi(s_{i-1}^{(t)},s_{i}^{(t)},s_{i+1}^{(t)})$. Such cellular automata are named elementary cellular automata or ECA \cite{wolfram02}. To specify a rule function $\phi$ a look up table can be given, for example as shown in Table \ref{tbl:rule}
\begin{table}[H]
\centering
\bigskip
\begin{tabular}{lll|l}
 $s_{i-1}^{(t)}$ & $s_{i}^{(t)}$ & $s_{i+1}^{(t)}$ & $s_{i}^{(t+1)}$ \\
 \hline
 1 & 1 & 1 & 0 \\
 1 & 1 & 0 & 0 \\
 1 & 0 & 1 & 1 \\
 1 & 0 & 0 & 0 \\
 0 & 1 & 1 & 1 \\
 0 & 1 & 0 & 1 \\
 0 & 0 & 1 & 1 \\
 0 & 0 & 0 & 0 \\
  \end{tabular}
  \caption{Look-up table for cellular automaton rule $46(00101110)$.}\label{tbl:rule}
\end{table}
When ordered lexicographically the rule has the binary representation $00101110$ equivalent to the decimal number $46$ (the reader should notice that according to the definition given in the introducion, the $\lambda$ parameter for this rule is $0.5$). CA rules are named by their decimal representation. The space of cellular automata with binary alphabet and nearest neighbor interaction will then be a discrete space of eight dimensions or an eight dimensional hypercube where there are $2^{2^3}=2^8=256$ rules. 

There is no apparent geometry in the CA discrete space, as there is no natural way to define a transition between CA rules in the hypercube allowing for the smooth change of one cellular automaton to a neighboring one while their behavior also changes gradually. One may think in ordering the CA rules as Gray codes \cite{savage97}, where successive CA differ by a change of only one entry in the rule table. However, Gray codes ordering does not lead to a corresponding smooth transition in the CA regimes. There is not even any apparent relation in behavior between adjacent rules. Consider rules $238$ and $110$ with binary representation $00101110$ and $01101110$, respectively. Both rules differ only in the seventh entry and yet, the former result always in a homogeneous state where each cell in the array has the same value, while the later, as already stated, can perform as a Universal Turing Machine. 

\begin{figure}[!ht]
\centering
\includegraphics*[scale=0.5]{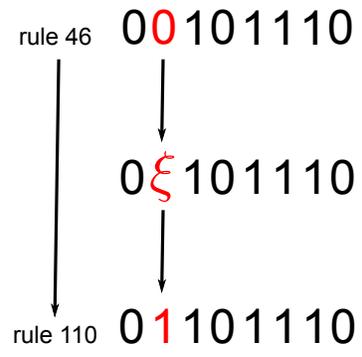}
\caption{Diagram of the continuous transition from rule $46$ to rule $110$ by changing the seventh entry.
}\label{fig:trans}
\end{figure}

The inability to define a natural ordering among the CA rules along which behavior changes gradually, has hindered the analysis of transitions between CA rules. To overcome this limitation, Pedersen devised a procedure to continuously deform  a CA rule into another at the expense of sacrificing the discrete nature of the cells states \cite{petersen90}. Consider rule $46$ and rule $110$ , a continuous transition from the former to the later can be done if we define a CA rule as $0\xi101110$, where $\xi$ goes from $0$ to $1$ in a continuous manner (Figure \ref{fig:trans}), such that for $\xi=0$ we get rule $46$ and for $\xi=1$ rule $110$ is recovered. In order for the transition to work, first a real valued function $\beta$ must be defined, for example $\beta(s_{i-1}s_{i}s_{i+1})=4 s_{i-1}+2 s_{i}+s_{i+1}$, next, an interpolating function $f(x)$ is used over the values of $\beta$, in our case: $f(\beta(s_{i-1}s_{i}s_{i+1})): [0,7] \rightarrow [0,1]$. The interpolating function is a real valued continuous function that must be consistent with the discrete case for integer values of its argument, but apart from that, it can be chosen in any way. Pedersen used three interpolating functions, one lineal, one quadratic and a third by trigonometric splines, finding that the results were qualitatively similar for every choice. We follow Pedersen and settle for the quadratic interpolating function 
\begin{equation}
 f(x)=\left \{ \begin{array}{ll}
 2[f(n+1)-f(n)](x-n)^2+f(n),& n \leq x \leq n+1/2\\\\
 2[f(n)-f(n+1)](x-n-1)^2+f(n+1),&n+1/2\leq x \leq n+1\end{array}\right.,
\end{equation}
where $n(\equiv\lfloor x \rfloor)$ is the largest integer smaller than $x$. As explained, if $n$ is an integer number $f(n)$ is given by the discrete CA rule.

The price paid in Pedersen extension is that the states of the cells are no longer binary values $\{0,1\}$ but any real value in the interval $[0,1]$. In order to recover the discrete binary nature of the cell states, we observe the system via a binarizing measuring device: we use the mean value over the whole cell array for a given time step as a threshold value, any cell with a state value over the mean is projected to $1$, otherwise to $0$.

In our simulations, an array of $5\times10^3$ cells were used and the initial random configuration (random binary digits were taken from www.random.org )  was left to evolve in $5\times 10^3$ time steps. Simulation results are then presented by averaging over ten simulations for each $\xi$ parameter value, each with different initial configurations. The last ten spatial configurations were taken for each run, so the final averaging was over $10^2$ spatial configurations. For testing the robustness of the result, calculations were occasionally performed for $10^4$, $5\times10^4$, $10^5$ and $10^6$ cells, and correspondingly equal time steps, showing no change in the results. Cyclic boundary conditions are used. In all cases, and contrary to Packard Monte Carlo simulations, the variance of the mean values for our simulations is too small even to be represented as error bars. 
 
Before we proceed, we need some measures that can capture the computational capabilities of a system and distinguish from the random portion of the output on the one hand, and data structuring on the other.

\section{Entropic measures} 

As a measure of unpredictability we use Shannon entropy rate $h_\mu$. When a bi-infinite sequence $S=\ldots s_{-3}s_{-2}s_{-1}s_{0}s_{1}s_{2}s_{3}\ldots$ of emitted symbols $s_i$ has been observed, $h_\mu$ is the amount of information produced per symbol. If we consider the bi-infinite sequence as a time series, then, it is the  amount of new information in an observation of let say cell $s_i$, considering the state of all previous cells $s_j$, $j<i$ \cite{crutchfield03}. In terms of Shannon entropy 
\begin{equation}
 h_{\mu}=H(s_i|\ldots, s_{i-1}),
\end{equation}
where $H(X|Y)$ denotes the Shannon conditional entropy of random variable $X$ given variable $Y$\cite{cover06}. 

The block entropy of a bi-infinite string $S$ measures the average uncertainty of finding blocks of a given length within the string. Block entropy is calculated by the equation
\begin{equation}
H_{S}(L)=-\sum\limits_{S^L \in \{ \mathcal{A}^L\}}p(S^L)\log p(S^L),
\end{equation}
where the sum goes over all sequences $S^L$ of length $L$ drawn from the alphabet $\mathcal{A}$, and $p(S^L)$ is the probability of observing one particular string $S^L$ in the bi-infinite string $S$. Entropy density, also known as entropy rate, can be understood as the unpredictability that remains when all that can be seen has been observed, and can also be defined through a limiting process of the normalized block entropies 
\begin{equation}
 h_\mu=\lim\limits_{L\rightarrow \infty}\frac{H_S(L)}{L}.\label{eq:hmu}
\end{equation}
For a necessarily finite data, the entropy rate has to be estimated. There is a body of literature dealing with the estimation of entropy rate for finite sequences \cite{schurmann99,rapp01,lesne09}. Here $h_\mu$  will be estimated through Lempel-Ziv factorization procedure that has been extensively discussed in the literature and has been used before in the context of CA \cite{kaspar87,estevez15}. In short, Lempel-Ziv factorization performs a sequential partition of a finite length string adding a new factor if and only if, it has not been found in the prefix up to the current position of analysis \cite{lz76}.  In order to emphasize that a finite size estimate of $h_\mu$ through the Lempel-Ziv procedure is being used, we change to the notation $h_{LZ}$ instead. 

Formally, if $s(i,j)$ is the substring of consecutive symbols $s_is_{i+1}\ldots s_j$ and $\pi$ the ''drop'' operator defined as $s(i,j)\pi=s(i,j-1)$ and consequently, $\pi^{l}=s(i,j-l)$, where $l$ is a positive number. The Lempel-Ziv factorization is a partition $F(S^N)$ of a finite string $S^N$ of length $N$
\begin{equation}
F(S^N)=s(1,l_1)s(l_1+1,l_{2})\dots s(l_{m-1}+1,N), \nonumber
\end{equation}
in $m$ factors such that each factor $s(l_{k-1}+1,l_k)$ complies with
\begin{enumerate}
 \item $s(l_{k-1}+1,l_k)\pi\subset s(1,l_k)\pi^2$
 \item $s(l_{k-1}+1,l_k)\not\subset s(1,l_k)\pi$ except, perhaps, for the last factor $s(l_{m-1}+1,N)$.
\end{enumerate}
The partition $F(S^N)$ is unique for every string. For example the sequence $u=101101110001$ has a Lempel-Ziv factorization $F(s)=1.0.11.0111.00.01$, each factor is delimited by a dot. For a data stream of length $1 \ll N < \infty$, $h_{LZ}$ is defined by 
\begin{equation}
 h_{LZ}(N)=\frac{C_{LZ}(S)}{N/\log{N}},\label{eq:hlz}
\end{equation}
 where $C_{LZ}(S)=|F(S^N)|$ is the number of factors in the Lempel-Ziv factorization. The entropy rate for an ergodic source is then given by\cite{ziv78}
\begin{equation}
 h_\mu=\limsup_{N\rightarrow\infty}h_{LZ}(N).
\end{equation}
For a finite data stream of length $N$ where the above limit can not be strictly realized, the entropy rate is then estimated by $h_{LZ}$ where we drop the argument $N$ when no ambiguity arises. The error in the estimation for a given value of $N$ has been discussed before\cite{amigo06}. 

Excess entropy $E$, introduced by Grassberger as effective complexity measure\cite{grassberger86}, has been used in several contexts \cite{crutchfield12} including the analysis of CA as a measure of correlation along different scales in a process and related to the intrinsic memory of a system. For a bi-infinite string, the excess entropy measures the mutual information between two infinite halves of the sequence and is related to the correlation of the symbols at all possible scales. Excess entropy can be defined in several equivalent ways, for example as the convergence of the entropy rate\cite{grassberger86,crutchfield03}
\begin{equation}
\begin{array}{ll}
 E(S) &=I[\ldots, s_{-1}: s_0, s_1, \ldots]\\
 &=\sum\limits_{L=1}^{\infty}[h_L(S)-h_\mu(S)]=\sum\limits_{L=1}^{\infty}\Delta H[L],\label{eq:excess}
 \end{array}
\end{equation}
 $I[X:Y]=H[X]+H[Y]-H[X,Y]$ is the mutual information between $X$ and $Y$, and $h_L(S)=H_S(L)-H_S(L-1)$, where $H(X)$ is the Shannon entropy of the random variable $X$. As already explained, mutual information is a measure of the amount of information one variable carries regarding the other, and it is a symmetric measure $I[X:Y]=I[Y:X]$. The information gain $\Delta H[L]=h_L(S)-h_\mu(S)$ measure how much the entropy density is being overestimated as a result of making measurements of the probability of only string blocks up to length $L$, in other words, how much information still has to be learned in order to assert the true entropy of the source\cite{crutchfield03}. As entropy decreases with length $L$, if such decrease is fast enough, the sum (\ref{eq:excess}) will converge. $E(S)$ is the intrinsic redundancy of the source or apparent memory of the source\cite{crutchfield03}. It measures structure in the sense of redundancy being a measure of pattern production.  Feldman\cite{crutchfield03} has proven that equation (\ref{eq:excess}) is equivalent, in the ergodic case, to the mutual information between two infinite halves of a bi-infinite string coming from the source output. Structure in this context then means how much information one halve carries about the other and vice-versa, a measure related again with pattern production and context preservation as redundancy. 

Numerically, when dealing with finite data streams, excess entropy has to be effectively estimated. Here we use a random shuffle based excess entropy estimate using the Lempel-Ziv estimation of entropy rate and given by \cite{melchert15}
\begin{equation}
 E_{LZ}=\sum\limits_{M=1}^{M_{max}}[h_{LZ}(S_{(M)})-h_{LZ}(S)].\label{eq:elz}
\end{equation}
$S_{(M)}$ is a surrogate string obtained by partitioning the string $S$ in non-overlapping blocks of length $M$ and performing a random shuffling of the blocks. The shuffling for a given block length $M$ destroys all correlations between symbols for lengths larger than $M$ while keeping the same symbol frequency. $M_{max}$ is chosen appropriately given the sequence length as to avoid fluctuations. In spite that $E_{LZ}$ is not strictly equivalent to the excess entropy as given by equation (\ref{eq:excess}), it is expected to behave in a similar manner \cite{melchert15}.

Finally, information distance $d(s,p)$ will be used. Information distance is derived from Kolmogorov randomness theory. The Kolmogorov randomness $s^{*}=K(s)$ of a sequence $s$, is the length of the shortest program ($s^{*}=|K(s)|$), that can reproduce the sequence $s$\cite{kolmogorov65}, accordingly, $K(s|p^{*})$, known as Kolmogorov conditional randomness, is the length of the shortest program able to reproduce $s$ if the program $p^*$, reproducing the string $p$ is given.  The information distance is defined by\cite{li04}

\begin{equation}
 d(s,p)=\frac{max\{K(s|p^{*}), K(p|s^{*})\}}{max\{K(s), K(p)\}}.\label{eq:dnid}
\end{equation}

$d(s,p)$ is an information based distance measuring, from an algorithmic perspective, how correlated are two sequences: if two sequences can be derived one from the other by a small-sized algorithm, then the corresponding $d(s,p)$ is small.

Following Estevez et. al.\cite{estevez15}, we estimate $d(s,p)$ via Lempel-Ziv by
\begin{equation}
 d_{LZ}(s,p)=\frac{C_{LZ}(sp)-min\{C_{LZ}(s), C_{LZ}(p)\}}{max\{C_{LZ}(s), C_{LZ}(p)\}}.\label{eq:dlz}
\end{equation}
which will have the same interpretation than $d(s,p)$  as much as the normalized (\ref{eq:hlz}) estimates the entropy density.

\section{Results}

\subsection{Continuously deformed cellular automata}

Having defined our entropic measures of randomness and complexity, we first explored the CA space. The logic followed was to focus in rule $110$ and all rules that result from flipping a single entry in its rule table (flipping can happen as $0\rightarrow 1$ or $0 \rightarrow 1$ depending on the rule entry), which leads to the eight possibilities given by rules $238$, $46$, $78$, $126$, $102$, $106$, $108$ and $111$. The $110$ rule is called the starting rule, and any of the eight rules related by a state flip in only one entry, the final or ending rule. We find in all cases that the transformation from rule $110$ to any of the ending rules proceeds, at a specific $\xi$ value, to a disordered state where entropy rate reaches the maximum value of $1$ and then at some other $\xi$ value, the system transforms to a behavior similar to the final rule. For specific ending rules, a clear jump in the excess entropy was found at the verge of falling into the final rule. That is the case for the transformation to ending rule $46$ (binary representation $00101110$), so the behavior around this rule was studied modifying each entry one at a time.

\begin{figure}[!ht]
\centering
\includegraphics*[scale=0.7]{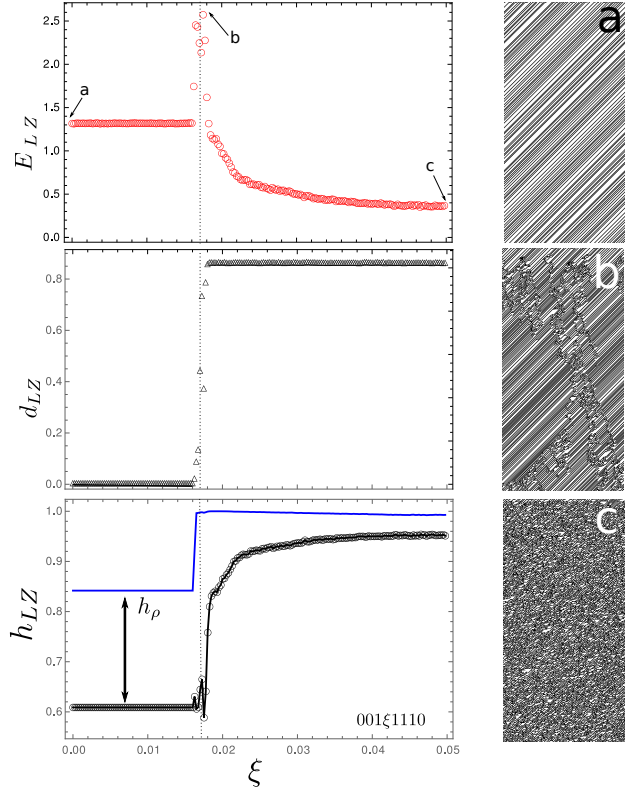}
\caption{ $E_{LZ}$ (upper left plot) and Lempel-Ziv information distance $d_{LZ}$ (middle left plot) for the cellular automaton rule $46$ ($\xi=0$) changing the fifth entry in the rule table towards rule $62$ ($\xi=1$). The plot shows $\xi$ values in the range $(0.00,0.05)$ to zoom into the initial phase transition region. Note the clear jump at around $\xi\approx 0.018$ for the entropy rate and the $d_{LZ}$ just before the $E_{LZ}$  jump reaches a maximum. The blue line labeled $h_{\rho}$ corresponds to the entropy rate of a uniform distributed random configuration of cells with the same density of one as the cellular automaton cells for the same $\xi$ value.  Calculations were performed with $5000$ cells for $5000$ time steps. The reported values are the average value over simulations for $10$ different random initial cell configurations. For each run, the last ten spatial configurations were taken making the effective averaging over $10^2$ spatial configurations. On the right, the spatiotemporal diagrams at the points marked by arrows in the upper left plot.
}\label{fig:ca46}
\end{figure}

Figure \ref{fig:ca46} shows the observed entropic measures for $\xi$ values near zero in the fifth entry ($001\xi1110$) of rule 46 ($\xi=0$) when moving towards rule $62$ ($\xi=1$). A small range of $\xi$ values is shown to emphasize details of the transition.  $E_{LZ}$ has a clear peak shaped maximum at a narrow window of $\xi$ values coincident with the jump in $h_{LZ}$. At the starting rule, $46$ the spatiotemporal diagram (Fig. \ref{fig:ca46} left {\bf{a}}) shows a shift diagram, the initial random conditions rapidly sets into a dynamic where the site values are spatially shifted with respect to the previous time step. At the value of $\xi=0.05$ (Fig. \ref{fig:ca46} left  {\bf{c}}) the spatiotemporal diagram is characteristic of cellular automata chaotic behavior as has been reported before (compare, for example, with the spatiotemporal diagram of rule 101 of class 3 in Wolfram automata classification \cite{wolfram02}). The spatiotemporal diagram, at the transformation point, when $\xi=0.018$ (Fig. \ref{fig:ca46} left  {\bf{b}}),  while largely keeping the shift behavior, shows the production of local structures that persist in time while traveling across the cell sites. Also, interactions between the structures are also seen.    The value of $\xi=0.018$, for which the peaking of $E_{LZ}$ is observed, we believe is a region of enhancement of computation capabilities of the system at the EOC. 

Large $E_{LZ}$ comes as a result of pattern prevalence over disorder: a high disorder configuration has a large entropy density estimated by $h_{LZ}$ and almost zero excess entropy estimated by $E_{LZ}$. This relation between entropy rate and excess entropy has been discussed numerous times in the past\cite{crutchfield03,feldman08,crutchfield12}. In a random sequence, there is no correlation between sites values at any length scale, and therefore, no mutual information can be found between two halves of the sequence. An increase in excess entropy can come from the formation of patterns not involving symbol erasure. If we relate computational capability, as stated in the introduction, as the storage, transmission, and manipulation of information, then the excess entropy increase signals an enhancement of such capabilities. Indeed, an increase in excess entropy means larger mutual information between two halves of a sequence which is directly related to storage and, less directly, to transmission of information. As excess entropy is also related to pattern formation, it can also signal the manipulation of information. So, the increase of $E_{LZ}$ is related to the enhancement of computation. The second condition that must be tested is that the jump in entropy density is witnessing a phase transition to a chaotic regime.

In order to test the hypothesis of transition to a chaotic regime, we performed the following simulation: the initial configuration was perturbed by changing one cell value in the cell array, and the system was left to evolve the same amount of time steps. We then calculated the Lempel-Ziv distance $d_{LZ}$ between the final configuration of the non perturbed and perturbed system. A high value of $d_{LZ}$ shows high sensitivity to initial conditions while the opposite is also true, low values of Lempel-Ziv distance is taken as showing low sensitivity to initial conditions. A similar procedure was used by Packard \cite{packard88} and Crutchfield et al. \cite{crutchfield93} using a different parameter based on Hamming distance and called difference-pattern spreading rate $\gamma$. As discussed above, $d_{LZ}$ differs from $\gamma$ in that the former is not Hamming distance related but instead, is based in how innovative is one sequence with respect to the other concerning pattern production \cite{estevez15}. In all our simulations, $d_{LZ}$ does not have the fluctuations reported for $\gamma$. In the two upper plot of figure \ref{fig:ca46} the excess entropy and the  $d_{LZ}$ as a function of the $\xi$ parameter are shown. At the same values of $\xi$ where the $E_{LZ}$ reaches for a maximum, the $d_{LZ}$ also jumps in a step-like manner. For the range of $\xi$ ($\in [0.0, 0.016]$) with low entropy rate, a low sensitivity to perturbation is shown by $d_{LZ}$. High sensitivity to initial conditions is a fingerprint of chaotic behavior. As soon as $h_{LZ}$ jumps, so does $d_{LZ}$, so the high entropy region is also a region with chaotic behavior. The region of peaked $E_{LZ}$  has intermediate values of entropy rate and $d_{LZ}$. 

Low values of entropy rate can come as a result of the erasing of one symbol, say $0$, at the expense of the other. In order to clarify how much of the entropy rate behavior could be due to the erasure (production) of $0$s ($1$s) the entropy rate for random arrangements of cell values, with different symbol density, was calculated for the whole range of possible values. In figure \ref{fig:ca46}b, the curve labeled $h_\rho$ is the entropy rate for the random arrangement having the same symbol density as the corresponding CA sequence with entropy density $h_{LZ}$ for the given $\xi$ value. The difference between both curves is a measure of how much lowering of the entropy rate is due to the structuring of the spatial cell values independent of the production of one symbol. Such a difference has been called multi-information in other contexts \cite{melchert15}. It must be noted that just before the entropy jump, in the region of the EOC, multi-information increases significantly as $h_\rho$ jumps before $h_{LZ}$. 

Besides the sensitivity to initial conditions, we looked for an additional criterion that points to the idea that after the $E_{LZ}$ peak, the system is indeed in chaotic behavior. A dynamical system is said to be chaotic if and only if it is transitive and its set of periodic points is dense in the phase space \cite{dubaq01}. For a system to be transitive, it is not hard to see that surjectivity is a necessary condition. Hedlund \cite{hedlund69} proved that a one dimensional discrete CA is surjective if and only if it is $k$-balanced for all $k \in \mathbb{N}$. We say that a discrete CA with radius $r=1$ is $k$-balanced if $\forall y \in S^{2(k-1)+1},\;|\{x\in S^{2k+1}| \phi(x)=y\}|=|S^{2}|$, where for a set $V$, $|V|$ means cardinality. For the binary alphabet used here, $1$-balance means that the rule table will have the same number of $1$s and $0$s, namely $4$. To prove $k$-balance, a random initial configuration is chosen, such that the density of either symbol is $1/2$, all spatial configurations up to time step $k$, must conserve the same density for the symbols. Rule $46$ is $1$-balanced but is not $k$-balanced for $k>1$. For continuous CA, proving $k$-balance for all $k \in \mathbb{N}$  can be more involved as time evolution occurs over a continuous state cell array, but the output is the discrete binary projection of the continuous cell array. We use the following criteria: if starting with a random initial binary configuration, after a sufficient number of time steps, the spatial configurations of the discrete projection preserves the density of both symbol at $1/2$, then the continuous rule is taken to be $k$-balanced.  If we assume that Hedlund condition is still necessary, we need to prove that the high entropy states in figure \ref{fig:ca46} are $k$-balanced, while the low entropy states are not. We performed such test numerically and found it to be true. The compliance of Hedlund rule, together with the sensitivity to the initial conditions points to the idea that certainly the values of $\xi$ where $E_{LZ}$ has a peaked maximum, is a region in the EOC. 

If $k$-balance of the CA is a necessary condition for chaotic behavior, then CA rules which are not $1$-balanced for more than one entry will never achieve such balance by changing a single entry in the rule table. Of the neighboring CA to rule $110$ that is the case for rules $238$, $126$ and $111$ where indeed we did not find any chaotic regime upon continuous deformation of any entry in the rule table and therefore no EOC enhancement (See supplementary material). Furthermore, it could be hypothesized that non-chaotic rules that are $1$-balanced or can be brought to $1$-balance by the deformation of a single entry in the rule table, are those were a transformation to a chaotic regime could be expected. Such is the case in the set of ECA for thirty-seven rules including rule $46$ but also rules $78$ and $108$ which are also neighbors of rule $110$. We have found a chaotic transition for rule $78$ at the rule table entry site corresponding to the $101$ triplet entry, and also enhancement at the EOC for this transition. Strong fluctuations in density were found in the EOC region, differing from rule $46$. The chaotic transition was also found for rule $108$ together with a smaller but clear peak of $E_{LZ}$ at the EOC. 

\subsection{Non-linear locally coupled oscillators (NLCO)}

Next, we set to explore if the same enhancement of computational capabilities at the EOC could be found in a system of non-linear coupled oscillators introduced by Alonso\cite{alonso17}. Alonso has argued about the similarities in the behavior of the  NLCO and the CA.  Coupled nonlinear oscillators have been extensively studied in relation with many natural phenomena such as chemical reactions, neural brain activity, insects synchronization \cite{mosekilde02}. The model used here consist in a system of locally coupled oscillators using, under the weak coupling assumption, a simply modified Adler equation \cite{adler73}. The system exhibits, for certain ranges of values of its control parameters, complex behavior. The local nature of the coupling, where the phase of one oscillator is directly related to two other neighbors, and the rich set of behaviors resembles that of the phenomena found in CA \cite{estevez18}. The array of coupled oscillators obeys, for the phase evolution, a system of equations given by
\begin{equation}
 \frac{d\theta_i}{dt}=\omega+\gamma\cos \theta_i+(-1)^i\left [\cos \theta_{i-1}+\cos \theta_{i+1}\right ].\label{eq:adler}
\end{equation}
$\omega$ is the proper frequency of the oscillators, and $\gamma$ controls the self-feedback of the oscillator relative to the coupling given by the third term in the right hand of equation (\ref{eq:adler}). The number of oscillators $N$ is taken even, and the alternating term in the coupling term is to guarantee balance in the interaction of the array of oscillators. Cyclic boundary conditions are enforced. The state of the system is given by the value, at a given time, of the phases $\theta_i$ of each oscillator which will be observed indirectly by the activity given by $\sin \theta_i$. As done with the CA, a similar binary partition scheme is introduced: the mean value of the activity is measured for a given time step and used as a threshold to binarize the data.

Simulations were performed following the same criteria as in the CA case, $5000$ coupled oscillators were left to evolve for $5000$ discrete time steps while the activity of the oscillators was recorded. Averaging was performed over $30$ different random initial phases configurations. The system of equations (\ref{eq:adler}) was solved numerically using a Runge-Kutta method of order $4$.  

\begin{figure}[!t]
\centering
\includegraphics*[scale=0.9]{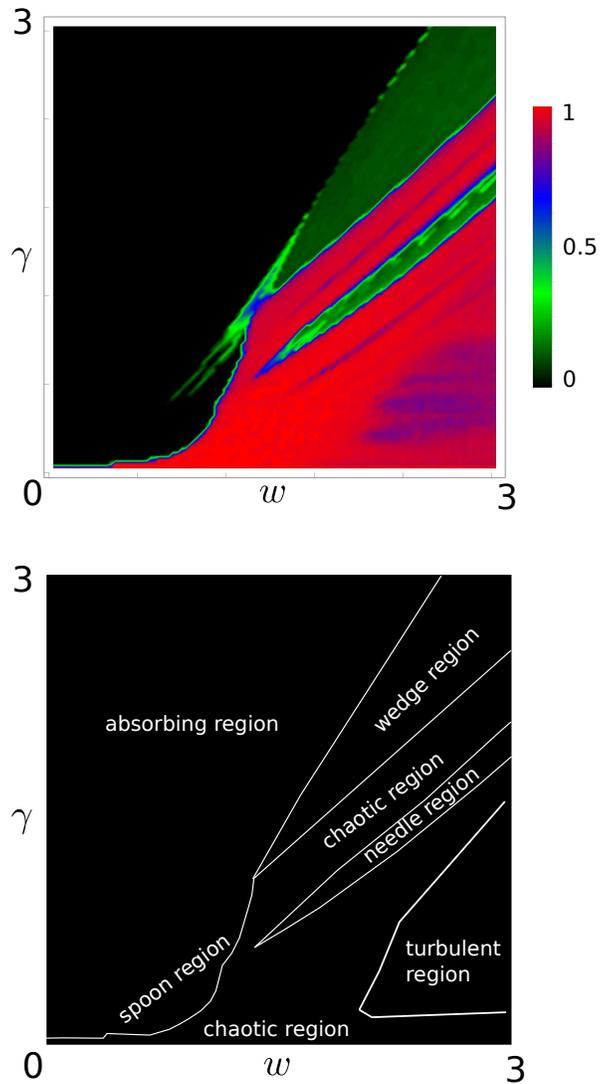}
\caption{The entropy density estimates for the NLCO from the Lempel-Ziv complexity over the control parameters $(\omega, \gamma)$ space. In all cases the number of oscillators is $N=10^4$ and $10^4$ time steps are taken. (abovel) Corresponds to the $c_{LZ}$ value from the final configuration of the oscillators after $10^4$ steps. Each point is the average of $100$ runs. (below) A diagram of the different regions identified in the two maps above. Taken from Ref. \cite{estevez18}.
}\label{fig:clz}
\end{figure}

\begin{figure}[!ht]
\centering
\includegraphics*[scale=0.5]{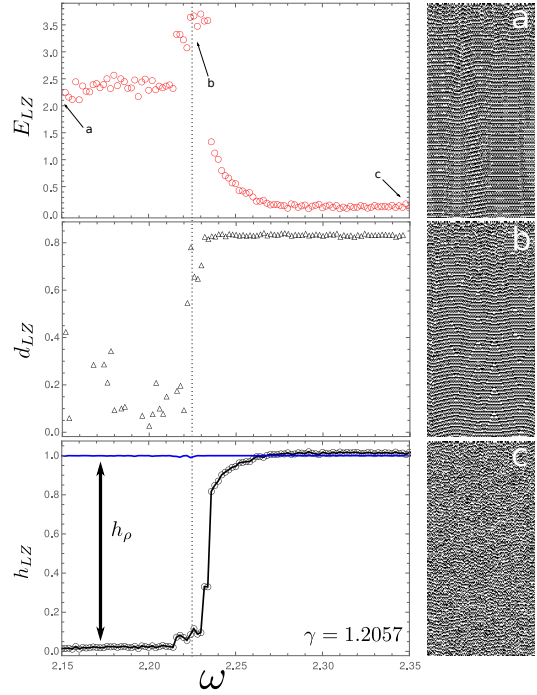}
\caption{The same measures as in Fig. \ref{fig:ca46} but for an array of $5000$ non-linear coupled locally coupled oscillators following an Adler type equation. A partition over the phase activity is taken using the mean value of the phases. Notation follows the ones used in Fig. \ref{fig:ca46}. In this case, $h_\rho$ is the entropy rate for the random arrangement having the same symbol density as the corresponding sequence of the NLCO for the given $\omega$ value. Again note the clear jump in entropy rate and $d_{LZ}$ around $\omega\approx 2.23$ just after $E_{LZ}$ reaches a maximum. Averaging was performed over $30$ random initial phase configurations for the oscillators. On the right, the spatiotemporal diagrams at the points marked by arrows in the upper left plot (High resolution spatiotemporal diagrams can be also found in the supplementary material).
}\label{fig:osc}
\end{figure}

In the case of the oscillators, the variables $(\omega, \gamma)$ vary continuously, and the control parameters space is therefore continuous. A detailed study of the ``phase'' diagram has been done before (Figure \ref{fig:clz} shows the phase diagram taken from Ref. \onlinecite{estevez18}). There are several particular regions in the control parameter space. We focused on the needle-shaped region surrounded by a connected chaotic region (Region labeled as needle in Figure \ref{fig:clz} below). Previous studies have indicated that the needle region exhibits low values of entropy rate and is insensitive to perturbations in the initial configuration of the oscillators, while the surrounding chaotic region has high entropy rate and high $d_{LZ}$ values (See figures 3 and 7 in Estevez et al. \cite{estevez18}). We chose a fixed $\gamma$ value of $1.2057$ and let $\omega$ vary between $2.10$ and $2.54$, enough to make a transition from the needle region to the chaotic region.  Figure \ref{fig:osc} shows the same type of plots as those shown in the CA case. Again a peaked shaped maximum of $E_{LZ}$ is witnessed at the same $\omega$ values where the entropy rate has a jump from low values to high disorder attaining the maximum possible value of $h_\mu=1$. The peaked maximum of $E_{LZ}$ also coincides with the jump in the $d_{LZ}$ values. The spatiotemporal diagrams at the transition point (Fig. \ref{fig:osc} left {\bf b} and supplementary material) shows  waves travelling across time and across the cell sites together with some local travelling structures, for smaller values of $\omega$ (Fig. \ref{fig:osc} left {\bf a} and supplementary material)  the local structures have disappeared and for larger values $\omega=2.35$ (Fig. \ref{fig:osc} left {\bf c} and supplementary material) a chaotic regime can be identified.   

In the case of the NLCO, no significant change in the production of one symbol is found, and the density of $1$s is more or less constant; therefore the decrease of $h_\mu$ for the smaller values of $\omega$ is a result of correlations between the oscillators phases. 

Similar to the CA, enhancement of computation at the EOC is also found.

\section{Complexity-entropy maps}

\begin{figure}[!ht]
\centering
\includegraphics*[scale=0.5]{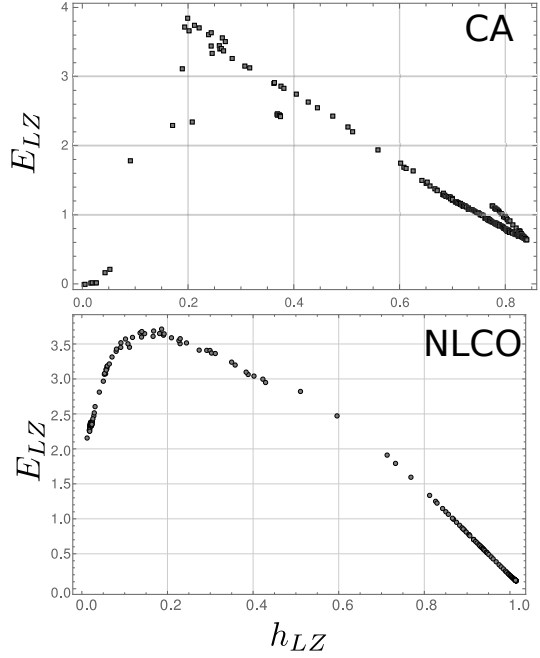}
\caption{Entropy-complexity diagrams for the cellular automaton rule 78 in the upper plot, and the system of non-linear oscillators in the lower plot. Rule $78$ was continuously deformed using as $\xi$ parameters the fifth entry in the rule table ($010\xi1110$). A binary generating partitions using the mean value as a threshold is used. Phase transition, in the case of rule $78$ shows at the diverging point  $h^{*}_{LZ}\approx 0.2$, below that point the regular behavior corresponding to rule $78$ prevails which behaves as a type II automata according to Wolfram classification. Above $h^{*}_{LZ}$ chaotic behavior is observed. The diagram corresponding to the oscillator system has a more smooth curve, but still, only a limited amount of randomness is allowed in the system in order for $E_{LZ}$ to achieve a maximum, above that level on unpredictability the system complexity start falling towards zero.  
}\label{fig:map}
\end{figure}

We finally look into the complexity-entropy map \cite{crutchfield12} of both systems. The entropy map for cellular automaton rule $78$ is shown in figure \ref{fig:map}a. Rule $46$ has a similar diagram.  The map was calculated by continuously varying the sixth entry in the rule table. Feldman et al. \cite{feldman08} discussed the interpretation of complexity-entropy diagrams; it measures the system intrinsic computation capabilities in a parameterless way. As already discussed, we also found enhanced computation at the EOC for rule $78$ (See supplementary material). The complexity-entropy map of figure \ref{fig:map}a is typical of systems transforming from non-chaotic to a chaotic regime, the reader may compare our plot with that reported by Crutchfield and Young for the logistic map exhibiting the well-known period-doubling cascading \cite{crutchfield89}. For values of entropy rate near zero, the system is rather simple, and the $E_{LZ}$ values are small. It is only with the introduction of some unpredictability that the system can accommodate more complex behavior. It is precisely at those value around $h_{LZ}=0.2$ where the $E_{LZ}$ attains a maximum that corresponds to the EOC. However, it is only to a certain value that a system can accommodate randomness while attaining higher intrinsic computation capability.

Further increase of entropy rate results in the decrease of the system intrinsic computation capability and at $h_{LZ}=1$, $E_{LZ}$ is again at zero. A similar interpretation can be made of the complexity-entropy diagram for the NLCO the main difference is that the curve is smoother pointing to a continuous first derivative. The striking similarity between our plots and those reported by Crutchfield and Young for well-studied systems, further points to the fact that the studied systems undergo a transition to chaotic behavior as the parameter $\xi$ is varied.

\section{Conclusions}

In conclusion, we have argued for the existence of enhanced computation at the edge of chaos for continuously deformed cellular automata, different from the reports of Langton and Packard. By using a continuously varying parameter that allows transforming from one CA rule to another, geometry in the CA space was recovered.   Our approach does not carry the limitations pointed out against Langton studies. The evidence heavily points to the existence, for specific CA rules, of a transition towards a chaotic regime where enhanced intrinsic computation can be found at the EOC. We directly measured such enhancement by looking into a measure of structural complexity while following the behavior of the entropy rate of the system. $k$-balanced analysis seems to be a useful tool to identify other rules where the same phenomena could be expected. The analysis was extended to a system of nonlinear coupled oscillators whose behavior has a strong resemblance to CA spatiotemporal patterns. For this system also improved intrinsic computation at the EOC was identified. The complexity-entropy diagrams show a striking resemblance to the ones reported for well studied non-linear systems were chaotic transitions are known to happen. 

\section{Acknowledgments}

This work was partially financed by FAPEMIG under the project BPV-00047-13. EER which to thank PVE/CAPES for financial support under the grant 1149-14-8. Infrastructure support was given under project FAPEMIG APQ-02256-12.
\bibliographystyle{unsrt}

\begin{thebibliography}{30}

\bibitem{crutchfield90}
J.~P. Crutchfield and K.~Young.
\newblock Computation at the onset of chaos.
\newblock In W.~H. Zurek, editor, {\em Complexity, entropy, and the physics of
  information}, pages 223--269. Addison Wesley, Redwood City, 1990.

\bibitem{langton90}
C.~G.Langton.
\newblock Computation at the edge of chaos.
\newblock {\em Physica D}, 42:12--37, 1990.

\bibitem{kauffman92}
S.~A. Kauffman and S.~Johnson.
\newblock Co-evolution to the edge of chaos: Coupled fitness landscapes, pised
  states, and co-evolutionary avalanches.
\newblock In C.G Langton, G.~Taylor, J.~Doyne Farmer, and S.~Rasmussen,
  editors, {\em Artificial Life II}, pages 325--368. Addison Wesley, Redwood
  City, 1992.

\bibitem{crutchfield93}
J.~P. Crutchfield, P.~T. Haber, and M.~Mitchell.
\newblock Revisiting the edge of chaos: evolving cellular automata to perform
  computations.
\newblock {\em Comput. Systems}, 7:89--130, 1993.

\bibitem{crutchfield12}
J.~P. Crutchfield.
\newblock Between order and chaos.
\newblock {\em Nature}, 8:17--24, 2012.

\bibitem{su89}
Z.~Su, R.~W. Rollins, and E.~R. Hunt.
\newblock Universal properties at the inset of chaos in diode resonator
  systems.
\newblock {\em Phys. Rev. A}, 40:2689--2697, 1989.

\bibitem{sole96}
R.~V. Sole, S.~C. Manrubia, B.~Luque, J.~Delgado, and J.~Bascompte.
\newblock Phase transitions and complex systems.
\newblock {\em Complexity}, XX:13--26, 1996.

\bibitem{bertschinger04}
N.~Bertschinger and T.~Natschlager.
\newblock Real time computation at the edge of chaos inrecurrent neural
  networks.
\newblock {\em Neural computation}, 16:1413--1436, 2004.

\bibitem{beggs08}
J.~M. Beggs.
\newblock The criticality hypothesis: how local cortical networks might
  optimiza information processing.
\newblock {\em Phil. Trans. of the Royal Soc. of London A: Math, Phys. and Eng.
  Science}, 366:329--343, 2008.

\bibitem{boedecker12}
J.~Boedecker, O.~Obst, J.~T. Lizier, N.~M. Mayer, and M.~Asada.
\newblock Information processing in echo state networks at the edge of chaos.
\newblock {\em Theory in Bioscience}, 131:205--213, 2012.

\bibitem{wolfram02}
S.~Wolfram.
\newblock {\em A new kind of science}.
\newblock Wolfram media Inc., Champaign, Illinois, 2002.

\bibitem{packard88}
N.~H.Packard.
\newblock Adaptation towards the edge of chaos.
\newblock In J.~A.~S. Kelso, A.~J. Mandell, and M.~F. Shlesinger, editors, {\em
  Dynamic patterns in complex systems}, pages 293--301. World Scientific,
  Singapore, 1988.

\bibitem{alonso17}
L.~M. Alonso.
\newblock Complex behavior in chains of nonlinear oscillators.
\newblock {\em Chaos}, DOI: 10.1063/1.4984800, 2017.

\bibitem{cook04}
M.~Cook.
\newblock Univrersality in elementary cellular automata.
\newblock {\em Complex Systems}, 15:1--40, 2004.

\bibitem{savage97}
C.~Savage.
\newblock A survey of combinatorial gray codes.
\newblock {\em SIAM Rev.}, 39:605--629, 1997.

\bibitem{petersen90}
J.~Pedersen.
\newblock Continous transitions of cellular automata.
\newblock {\em Complex systems}, 4:653--665, 1990.

\bibitem{crutchfield03}
J.~P. Crutchfield and D.~Feldman.
\newblock Regularities unseen, randomness observed: Levels of entropy
  convergence.
\newblock {\em Chaos}, 13:25--54, 2003.

\bibitem{cover06}
T.~M. Cover and J.~A. Thomas.
\newblock {\em Elements of information theory. Second edition}.
\newblock Wiley Interscience, New Jersey, 2006.

\bibitem{schurmann99}
T.~Schurmann and P.~Grassberg.
\newblock Entropy estimation of symbol sequence.
\newblock {\em Chaos}, 6:414--427, 1999.

\bibitem{rapp01}
P.~E. Rapp, C.~J. Cellucci, K.~E. Korslund, T.~A.~A. Watanabe, and M.~A.
  Jimenez-Montano.
\newblock Effective normalization of complexity measurements for epoch length
  and sampling frequency.
\newblock {\em Phys. Rev. E}, 64:016209--016217, 2001.

\bibitem{lesne09}
A.~Lesne, J.L.Blanc, and L.~Pezard.
\newblock Entropy estimation of very short symbolic sequences.
\newblock {\em Phys. Rev. E}, 79:046208--046217, 2009.

\bibitem{kaspar87}
F.~Kaspar and H.~G. Schuster.
\newblock Easily calculable measure for the complexity of spatiotemporal
  patterns.
\newblock {\em Phys. Rev. A}, 36:842--848, 1987.

\bibitem{estevez15}
E.~Estevez-Rams, R.~Lora-Serrano, C.~A.~J. Nunes, and B.~Arag\'on-Fern\'andez.
\newblock {L}empel-{Z}iv complexity analysis of one dimensional cellular
  automata.
\newblock {\em Chaos}, 25:123106--123116, 2015.

\bibitem{lz76}
A.~Lempel and J.~Ziv.
\newblock On the complexity of finite sequences.
\newblock {\em IEEE Trans. Inf. Th.}, IT-22:75--81, 1976.

\bibitem{ziv78}
J.~Ziv.
\newblock Coding theorems for individual sequences.
\newblock {\em IEEE Trans. Inf. Th.}, IT-24:405--412, 1978.

\bibitem{amigo06}
J.~M. Amigo and M.~B. Kennel.
\newblock Variance estimators for the lempel ziv entropy rate estimator.
\newblock {\em Chaos}, 16:43102--43108, 2006.

\bibitem{grassberger86}
P.~Grassberger.
\newblock Towards a quantitative theory of self-generated complexity.
\newblock {\em Int. J. Theo. Phys.}, 25:907--938, 1986.

\bibitem{melchert15}
O.~Melchert and A.~K. Hartmann.
\newblock Analysis of the phase transition in the two-dimensional ising
  ferromagnet using a lempel-ziv string-parsing scheme and black-box
  data-compression utilities.
\newblock {\em Phys. Rev. E}, 91:023306--023317, 2015.

\bibitem{kolmogorov65}
A.~N. Kolmogorov.
\newblock Three approaches to the concept of the amount of information.
\newblock {\em Probl. Inf. Transm. (English Trans.).}, 1:1--7, 1965.

\bibitem{li04}
M.~Li, X.~Chen, X.~Li, B.~Ma, and P.~M.~B. Vitanyi.
\newblock The similarity metric.
\newblock {\em IEEE Trans. Inf. Th.}, 50:3250--3264, 2004.

\bibitem{feldman08}
D.~P. Feldman, C.~S. McTeque, and J.~P. Crutchfield.
\newblock The organization of intrinsic computation: complexity-entropy
  diagrams and the diversity of natural information processing.
\newblock {\em Chaos}, 18:043106--043121, 2008.

\bibitem{dubaq01}
J.-C Dubacq, B.~Durand, and E.~Formenti.
\newblock Kolmogorov complexity and cellular automata classification.
\newblock {\em Th. Comp. Science}, 259:271--285, 2001.

\bibitem{hedlund69}
G.~A. Hedlund.
\newblock Endomorphism and automorphism of the shift dynamical systems.
\newblock {\em Math. Systems Theory}, 3:320--375, 1969.

\bibitem{mosekilde02}
E.~Mosekilde, Y.~MAistrenko, and D.~Postnov.
\newblock {\em Chaotic Synchronization: Application to Living Systems}.
\newblock World Scientific, Singapore, 2006.

\bibitem{adler73}
R.~Adler.
\newblock A study of locking phenomena in oscillators.
\newblock {\em Proc. of the IEEE}, 61:1380--1385, 1973.

\bibitem{estevez18}
E.~Estevez-Rams, D.~Estevez-Moya, and B.~Aragon-Fernandez.
\newblock Phenomenology of coupled nonlinear oscillators.
\newblock {\em Chaos}, 28:23110--23121, 2018.

\bibitem{crutchfield89}
J.~P. Crutchfield and K.~Young.
\newblock Inferring statistical complexity.
\newblock {\em Phys. Rev. Lett}, 63:105--108, 1989.

\end{thebibliography}

\pagebreak

\pagebreak

\end{document}